\documentclass[twocolumn]{emulateapj}

\shorttitle{}
\shortauthors{Ben\'{i}tez-Llambay et al.}

\begin{document}

\title{Dwarf Galaxies and the Cosmic Web}

\author{Alejandro Ben\'{i}tez-Llambay\altaffilmark{1}, Julio F. Navarro\altaffilmark{2}, Mario G. Abadi\altaffilmark{1}, %
Stefan Gottl\"ober\altaffilmark{3}, Gustavo Yepes\altaffilmark{4}, Yehuda Hoffman\altaffilmark{5}, Matthias Steinmetz\altaffilmark{3}
}

\altaffiltext{1}{Observatorio Astron\'omico, Universidad Nacional de C\'ordoba, C\'ordoba, X5000BGR, Argentina}
\altaffiltext{2}{Department of Physics \& Astronomy, University of Victoria, Victoria, BC, V8P 5C2, Canada}
\altaffiltext{3}{Leibniz Institute for Astrophysics, An der Sternwarte 16, 14482 Potsdam, Germany}
\altaffiltext{4}{Departamento de F\'isica Te\'orica, Universidad Aut\'onoma de Madrid, 28049, Madrid, Spain}
\altaffiltext{5}{Racah Institute of Physics, The Hebrew University of Jerusalem, Jerusalem, 91904, Israel}

\begin{abstract}
  We use a cosmological simulation of the formation of the Local Group of Galaxies to identify a mechanism that enables the removal of baryons from low-mass halos without appealing to feedback or reionization.  As the Local Group forms, matter bound to it develops a network of filaments and pancakes. This moving web of gas and dark matter drifts and sweeps a large volume, overtaking many halos in the process. The dark matter content of these halos is unaffected but their gas can be efficiently removed by ram-pressure. The loss of gas is especially pronounced in low-mass halos due to their lower binding energy and has a dramatic effect on the star formation history of affected systems. This ``cosmic web stripping'' may help to explain the scarcity of dwarf galaxies compared with the numerous low-mass halos expected in $\Lambda$CDM and the large diversity of star formation histories and morphologies characteristic of faint galaxies. Although our results are based on a single high-resolution simulation, it is 
likely that the hydrodynamical interaction of dwarf galaxies with the cosmic web is a crucial ingredient so far missing from galaxy formation models.
  \end{abstract}

\section{Introduction}
A long-standing puzzle concerns the striking difference between the
shape of the galaxy stellar mass function and the cold dark matter
halo mass function on dwarf galaxy scales~\citep{White1978}: dwarfs are
much less numerous than halos massive enough, in principle, to host
them~\citep{Klypin1999,Moore1999}. This is usually reconciled by
appealing to baryonic processes that reduce drastically the efficiency
of galaxy formation in low-mass halos. In order to match the latest
galaxy clustering and abundance data, recent
models~\citep{Moster2010,Guo2010} require that very few galaxies must
form in halos with virial\footnote{Virial quantities are measured
  within spheres of mean density 200 times the critical density for
  closure, and are identified with the subscript ``200''.}  mass below
$\sim 10^{10}\, M_\odot$. In addition, these models suggest a steep
relation between dwarf galaxy mass and halo mass, so that most dwarfs
(defined in terms of their stellar mass; $10^6 < M_{\rm gal}/M_\odot <
10^{9.5}$) are expected to form in halos spanning a narrow range in
virial mass.

\begin{figure*}[!t]
\begin{center}
\epsscale{1.0}
\plotone{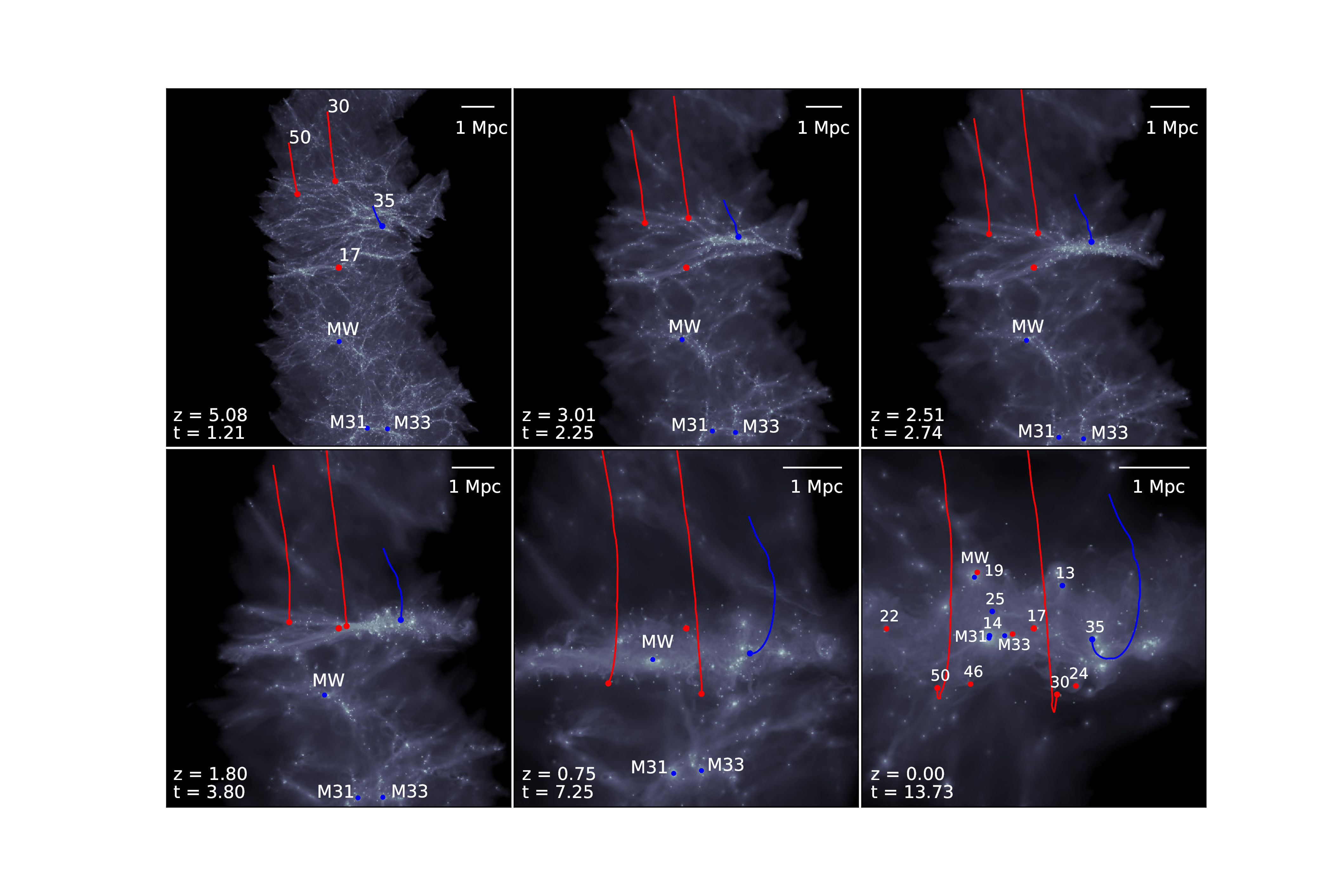}
\end{center}
\caption{Gas distribution in the high-resolution region at various times.
The positions of the three most massive systems, analogous to the main spiral galaxies of the Local Group, are labelled ``MW'', ``M31'', and ``M33'', respectively.  The central galaxies of halos with virial mass exceeding $10^{10}\, M_\odot$ and within $\sim 1.8$ Mpc from the Local Group barycenter at $z=0$ are numbered in the final snapshot. The solid curves track the location of a few low-mass halos, chosen to illustrate the interaction between them and the ``pancake'' that forms at $z\sim 2$. Tracks begin at $z=41$ and end at the current position of each galaxy. \label{FigA}}
\end{figure*}

These scalings are thought to arise from the combined effects of photoionization by the cosmic UV background~\citep[see, e.g.,][]{Bullock2000} and feedback energy from evolving stars ~\citep[e.g.,][]{Benson2003}. However, there are indications that these mechanisms alone may fall short. Ionizing radiation can effectively prevent gas accretion only in halos below $10^9\, M_\odot$~\citep{Crain2007}; in addition, the feedback energy required to regulate star formation seems to exceed what is released by supernovae~\citep[see, e.g.,][and references within]{Scannapieco2011,Brook2012}, a difficulty exacerbated by the low efficiency with which feedback couples to the surrounding medium. This is especially true for the faintest galaxies, where the small number of stars excludes feedback as a major energetic source, unless ad-hoc modifications to the IMF are invoked~\citep{Penarrubia2012}. 

Finally, there is growing observational evidence that many dwarf galaxies form in halos with masses well below $10^{10} \, M_\odot$~\citep{Boylan-Kolchin2011,Ferrero2012}, raising concerns about the general validity of the assumption that dwarfs inhabit relatively massive halos. Identifying an additional mechanism able to reduce the efficiency of galaxy formation in low-mass halos would help to allay concerns that modifications to the current paradigm (such as a ``warm dark matter'' particle) might be needed in order to reconcile $\Lambda$CDM with observation.

\section{The CLUES Local Group simulation}

We address these issues here using a Local Group simulation from the CLUES\footnote{Constrained Local UniversE Simulations: {\tt http://www.clues-project.org}} Project~\citep{Gottloeber2010}. The simulation follows the environment of three dark matter halos that match the relative positions of the three main spirals in the Local Group: the Milky Way (MW), Andromeda (M31) and Triangulum (M33) galaxies. These halos are identified in a cosmological box $87.7$ Mpc on a side, constrained by observational data to match our nearby large-scale environment. 

The Local Group region is then resimulated using a zoom-in technique~\citep{Klypin2001} where an 80 Mpc$^3$ volume is evolved at high-resolution embedded in a lower-resolution version of the full box. The high-resolution region (a roughly spherical region of $2.74$ Mpc radius at $z=0$) is filled with $\sim 53$ million dark matter particles of mass $3.48 \times 10^5 \, M_\odot$ and with the same number of $6.05 \times 10^4 \, M_\odot$ gas particles.  The Plummer-equivalent gravitational softening is $\epsilon = 0.14$ kpc. Stars are modeled as collisionless particles of mass $m_{\rm str}= 3.03 \times 10^4 \, M_{\odot}$ spawned by gas particles when they are eligible to form stars.

The simulation adopts cosmological parameters consistent with WMAP-3~\citep{Spergel2007} and were run with {\small GADGET-2}~\citep{Springel2005} code. They include an evolving cosmic ionizing UV background~\citep{Haardt1996}, star formation, supernova feedback and isotropic winds according to~\citep{Springel2003}; all is calibrated to reproduce, approximately, the stellar masses and morphologies of the three main Local Group spirals.  We refer the interested reader to \citet{Gottloeber2010} for further details about these simulations.

\begin{figure*}
\begin{center}
\epsscale{0.8}
\plotone{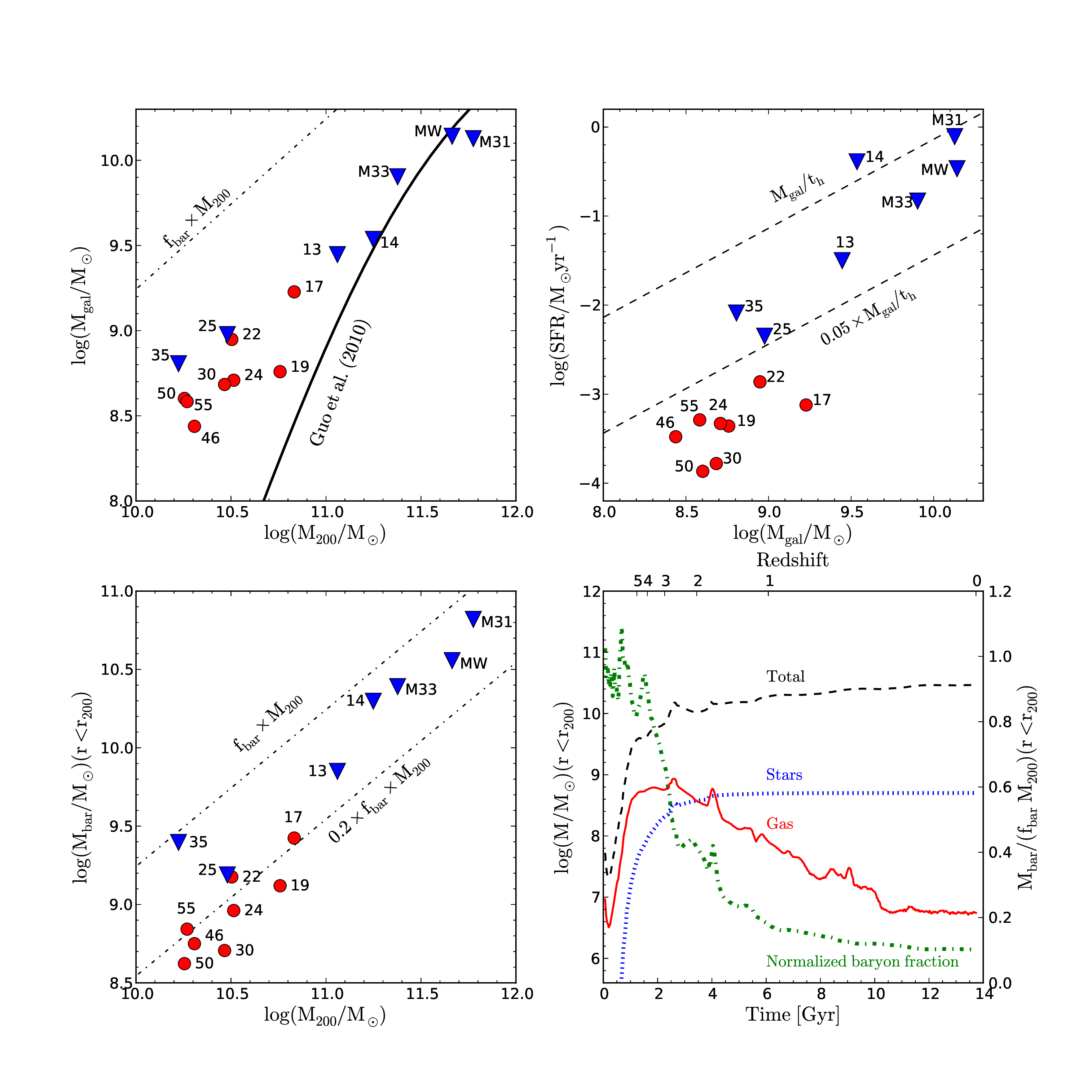}
\end{center}
\caption{Stellar mass within the galactic radius (top left) and total baryonic mass within the virial radius (bottom left) as a function of halo virial mass. The top-right panel shows the present-day star formation rate as a function of galaxy stellar mass.
Two populations are identified: (i) galaxies that form stars at rates comparable to their past average (blue triangles), and (ii) those where star formation has largely ceased (red circles). Galaxies in the latter group have lost more than $\sim 80\%$ of their available baryons. At given halo mass, these galaxies have formed substantially fewer stars than their star-forming counterparts, suggesting that the loss of baryons and lack of star formation are not due to feedback from evolving stars. The evolution of the 
mass within the virial radius of one such galaxy (galaxy ``30'') illustrates the sudden loss of baryons that results from ram-pressure arising from crossing a large-scale pancake at $z\sim 2$ (bottom-right). The virial mass more than doubles thereafter, but the total baryonic mass barely grows; at $z=0$ the baryon content of this halo is less than $10\%$ that expected from the universal ratio. \label{FigB}}
\end{figure*}

\section{RESULTS}

Fig.~\ref{FigA} shows the evolution of the gas component of the high-resolution region. Labels identify the position of the three main halos, where the galaxy analogs to MW, M31, and M33 form. Several other galaxies form in halos of lower mass, as well as in subhalos of the main three spirals; we shall exclude from the analysis that follows all satellites and focus on the central galaxies of halos with $M_{200}>10^{10}\, M_\odot$. This choice ensures that the properties we study are not influenced by effects, such as tidal stripping, that arise from the interaction of a galaxy with the halo of a more massive system.

The evolution of the gaseous component depicted in Fig.~\ref{FigA} is similar to that of the dark matter (not shown) when viewed on these large scales, and shows the early development of highly aspherical features such as filaments and pancakes that characterize structure formation in $\Lambda$CDM. The tracks that crisscross the figure highlight the trajectories of several dwarf galaxies and illustrate the paths they follow as they are assembled into the Local Group.
One of the galaxies (identified as ``17'') is at rest at the origin of the reference frame.

Fig.~\ref{FigB} summarizes the properties of these galaxies\footnote{We compute ``galaxy properties'' within the galactic radius, defined by $r_{\rm gal}=0.15\, r_{200}$.}  at $z=0$. The top-right panel shows that the simulated galaxies form two groups, one of systems that are forming stars at rates roughly comparable to their past average (blue triangles) and a second group where star formation has largely ceased (red circles). The latter are galaxies that have lost most of their baryons during their evolution; indeed, none of these galaxies has retained more than $\sim 20\%$ of the baryons within their virial radius, (bottom-left panel). It might be tempting to ascribe the loss of baryons to the effects of feedback-driven winds but there are galaxies that have retained more baryons (bottom-left panel) and where star formation has continued to the present (top-right panel), despite having formed more stars overall and inhabiting halos of similar virial mass.

A clue to the origin of the baryon-poor, non-star forming galaxies is provided in the bottom-right panel of the same figure. This shows the evolution of the mass within the virial radius of the main progenitor of galaxy $30$. The solid and dotted curves indicate the mass in the gaseous and stellar components (the dashed curve indicates the total; scale on left) whereas the dot-dashed green curve indicates the baryon fraction of the system, $M_{\rm bar}/(f_{\rm bar}\, M_{200})$ (scale on right), where $M_{\rm bar}$ is the sum of gas and stellar masses and $f_{\rm bar}=\Omega_b/\Omega_{\rm M}$ is the universal baryon fraction. This shows that most baryons are lost over a short period of time, between $z=3$ and $z=2$. Very few stars form after $z=3$, so that baryon loss cannot be due solely to feedback-driven winds.

\begin{figure}
\begin{center}
\epsscale{1.2}
\plotone{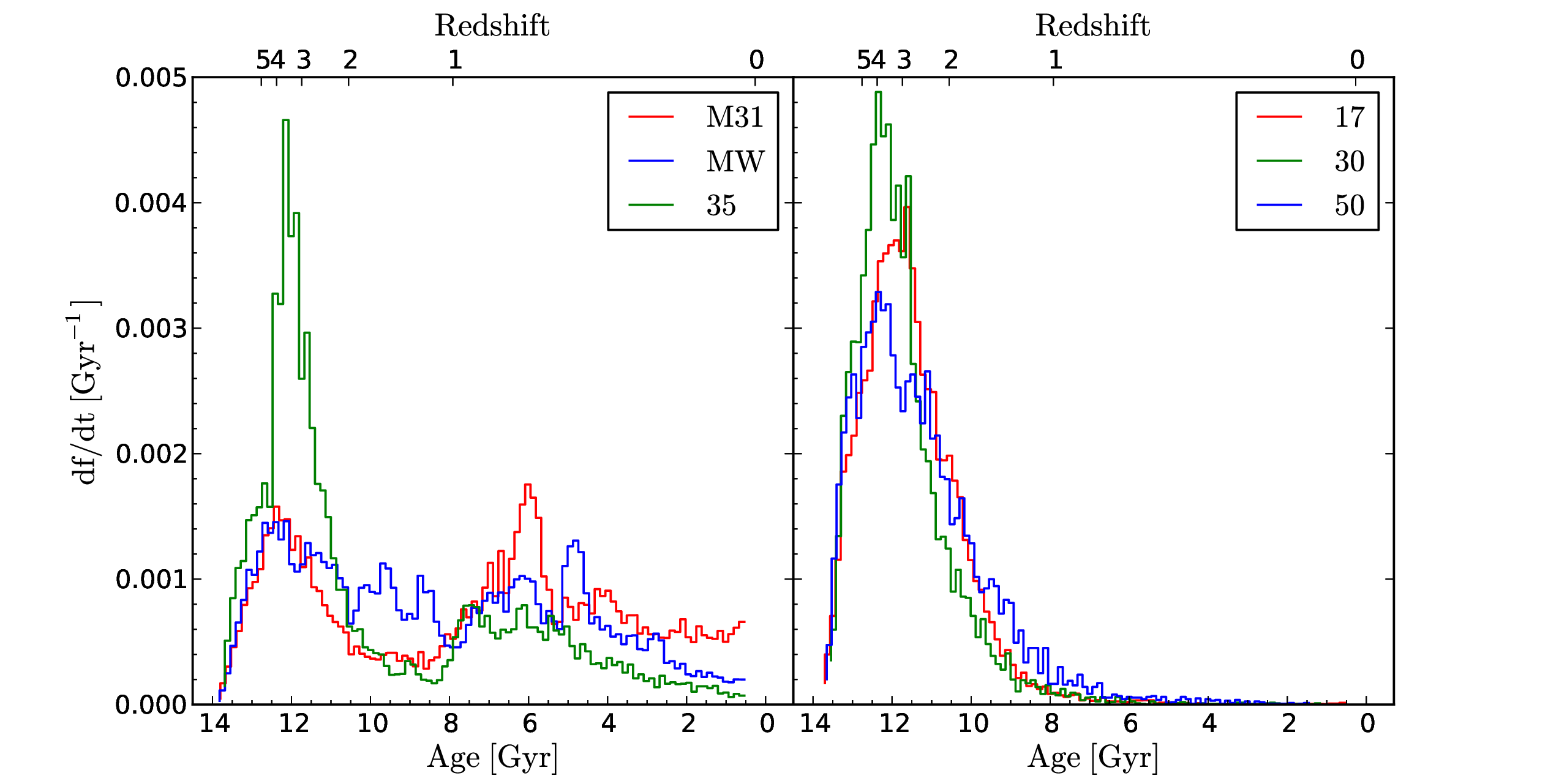}
\end{center}
\caption{Star formation history of galaxies affected by cosmic web stripping of their gaseous envelopes (right-hand panel). The histograms there show the distribution of stellar ages of three galaxies whose trajectories are tracked in Fig.~\ref{FigA}. All of these systems traverse a large-scale ``pancake'' at roughly the same time, $z\sim 2$, when they are stripped of most of their gas. Without this reservoir, star formation ceases soon thereafter, once the dense gas that remains is transformed into stars. By
  contrast, star formation in halos that are not affected by cosmic web stripping is able to continue until the present. These are galaxies that do not cross any major filament during their evolution, or that are massive enough to resist stripping. Three examples are shown in the left-hand panel. \label{FigC}}
\end{figure}

The loss of baryons highlighted above is the main reason why this galaxy has stopped forming stars. Indeed, star formation ceases in galaxy $30$ just after $z\sim 2$, as may be seen in the right-hand panels of Fig.~\ref{FigC}, where we show the distribution of stellar ages of galaxies $30$, $50$ and $17$. All three galaxies have similar star formation histories, and stop forming stars at roughly the same time. Their star formation histories differ markedly from those of galaxies that are still forming stars at $z=0$, three of which are shown in the left-hand panel of Fig.~\ref{FigC}.

What causes the peculiar star formation history of galaxy 30, which is in turn paralleled by galaxies 17 and 50? It is not their halo mass; for example, the virial mass of galaxy 35 is similar to that of galaxy 50, but is still forming stars at $z=0$ at a rate roughly $50$ times higher. We can gain some insight by considering in more detail the trajectories of galaxies 17, 30, and 50 shown in Fig.~\ref{FigA}.  This figure reveals a common feature in the evolution of these three galaxies: they all cross more or less simultaneously a large pancake of gas at $z\sim 2$. In the reference frame chosen for that plot (where galaxy 17 is at rest), the pancake sweeps past galaxy 17 at $z \sim 1.8$; galaxies 30 and 50 overtake the pancake at about the same time.

\begin{figure*}
\begin{center}
\epsscale{1.0}
\plotone{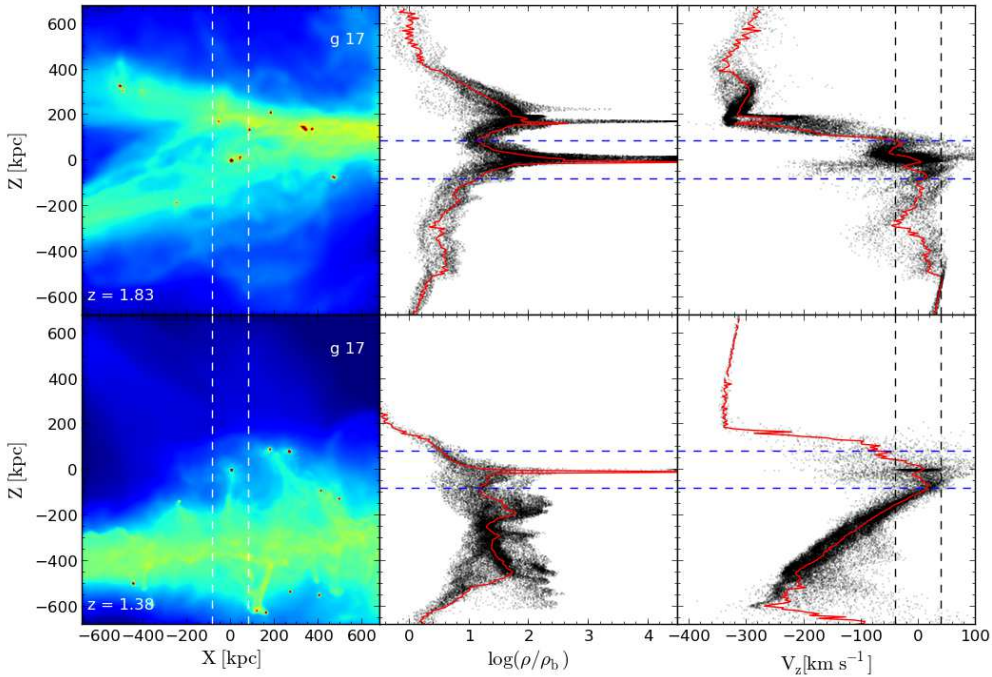}
\end{center}
\caption{Details of the ram-pressure stripping of galaxy 17. 
Left panels show the projected gas distribution in a cube $1.2$ comoving Mpc on a side, centered on the galaxy. The projection is identical to that of Fig.~\ref{FigA} and has been chosen so that the ``pancake'' that forms early on is approximately perpendicular to the $z$ axis. Colors indicate gas density in a logarithmic scale, normalized to the mean baryonic density of the Universe at that time, $\rho_b(z)=\Omega_b \, \rho_{\rm crit}(z)$. The middle panels show the gas density along the narrow vertical cylinder indicated by the dashed lines in the left panels. Right-hand panels indicate the $z$-component of the gas velocity along the cylinder in the rest-frame of the galaxy.  The vertical dashed lines in the right-hand panels indicate the current virial velocity of galaxy 17. Each row corresponds to a different time, chosen to illustrate just before and just after galaxy 17 plows through the pancake, 
losing most of its gas. Although the pancake is less dense than the galaxy, their relative velocity is nearly ten times the virial velocity of the galaxy. The resulting ram-pressure strips most of the gas off the galaxy, leaving behind ``streams'' of gas that are effectively lost.
\label{FigD}}
\end{figure*}

A more detailed analysis of the galaxy-pancake interaction is presented in Fig.~\ref{FigD}, where we show, at two different times (just before and after the interaction), the density and velocity profiles of galaxy 17 and the pancake. The two panels on the left show the gas distribution (colored by density on a logarithmic scale) in the vicinity of the galaxy, which is at rest at the coordinate origin. The pancake is moving along the $z$ axis, from positive-$z$ to negative-$z$ values. The two vertical dashed lines indicate the virial radius of galaxy 17 at the current time. 

At $z=1.83$, corresponding to the top row, the pancake is still above the galaxy, but approaching at high speed. This may be seen in the right-hand panels, which show the $z$-component of the gas velocity along the cylinder delineated by the vertical lines in the left-hand panel. The cylinder, whose radius is chosen to be the same as the
virial radius of the galaxy, is aligned with the $z$ axis and therefore roughly perpendicular to the pancake. The gas density in the cylinder, as a function of $z$, is shown in the middle panels. At $z=1.38$, which corresponds to the bottom row, the pancake has already swept past the galaxy and is found at $z\sim -400$ kpc.

It is clear from Fig.~\ref{FigD} that the interaction between the galaxy and the pancake leads to the removal of much of the gaseous halo of galaxy 17, which may be seen streaming down towards the pancake in the bottom-left panel. This is clear evidence that the ram-pressure of the interaction is responsible for the sudden loss of
baryons that affects galaxy 17 at $z\sim 1.8$. Since not all of the gas is removed, the galaxy is able to continue forming stars for a little longer before running out of gas. Although not shown in Fig.~\ref{FigD}, a similar process affects galaxies 30 and 50 as they traverse the pancake, as well as the rest of the galaxies labeled in Fig.~\ref{FigB} as non-star forming. 

One may wonder how the pancake, which is only a few times overdense relative to the mean baryonic density of the Universe (see middle panels of Fig.~\ref{FigD}), is able to remove the much denser gas that fills the halo of galaxy 17. The main reason for this is the high speed at which the galaxy plows through the pancake. The effective ram-pressure on the galaxy is proportional to the density of the pancake times the velocity of the interaction squared, $P_{\rm ram}\propto \rho_{\rm p} V_{\rm p}^2$, where $V_{\rm p}$ is the velocity of the pancake if the galaxy is assumed at rest. This pressure must exceed that exerted on the gas by the restoring force of the halo, which is roughly proportional to $P_{\rm gal}\propto \rho_{\rm gal} V_{200}^2$, where $\rho_{\rm gal}$ is the gas density and $V_{200}$ is the halo virial velocity.

For the case illustrated in Fig.~\ref{FigD}, the pancake velocity is of order $300$ km/s, almost an order of magnitude higher than the virial velocity of the halo at the time, $V_{200} \sim 40$ km/s. This means that the pancake can strip gas from the galaxy even if the galactic gas is $\sim (V_{\rm p}/V_{200})^2\sim 60$ times denser. A further condition must be satisifed for ram-pressure stripping to be effective, namely that the pancake is massive enough so that the column density of gas that sweeps past the galaxy is comparable to its total gas content. We have verified that this is indeed the case: the total pancake gas mass intersected by the cylinder shown in Fig.~\ref{FigD} exceeds the gas content of galaxy 17 by more than a factor of two.

\section{Discussion and Conclusions}
\label{SecConc}

Although gas removal through interaction with the cosmic web is a robust feature of the simulation, the fate of the gas that remains in the galaxy or is accreted later is less clear, since it is subject to the complex (and poorly-modeled) process of star formation and associated feedback.  Indeed, there is indication that the star formation recipes adopted in CLUES might need revision; note, for example, that most dwarfs lie above the $M_{\rm gal}$-$M_{200}$ relation indicated by recent abundance-matching analysis~\citep{Guo2010} (top-left panel of Fig.~\ref{FigB}). In addition, there seem to be fewer non-star-forming isolated dwarfs in the local Universe than in our simulation~\citep{Geha2012}. Therefore, the actual stellar masses of the simulated dwarfs or the lack of ongoing star formation should not be necessarily considered a defining characteristic of the scenario we describe here.  Our results should be interpreted as emphasizing the importance of the cosmic web on the overall baryonic content of 
dwarf galaxy halos rather than as definitive predictions to be compared directly with observation.

The importance of ``cosmic web stripping'' has not been recognized before because it is a purely hydrodynamical effect that would be completely missed in semi-analytic modeling and that requires simulations of large volumes able to resolve properly both the cosmic web and the internal properties of low-mass halos. This demands
numerical resources that are only now becoming available. It also requires careful tracking of the baryonic content of dwarf galaxies in order to discriminate between the competing effects of re-ionization, feedback, and stripping. Since all of these effects act to remove baryons from low-mass halos, it is even possible that the effects of
cosmic web stripping may have already been seen~\citep{Sawala2012} but misidentified. 

Cosmic web stripping is a natural consequence of the shape of the $\Lambda$CDM power spectrum on dwarf galaxy scales.  It results from the nearly simultaneous assembly of very different mass scales; at the same time that the $10^{10}\, M_\odot$ halos are being assembled, the collapse of the nearly $10^{13}\, M_\odot$ Local Group has already
started and is well under way in one dimension, leading to the formation of a large pancake. Although the CLUES simulation we analyze here is too small to enable a proper assessment of the importance of cosmic web stripping outside Local Group-like environments, we see no reason why this mechanism should not affect a substantial fraction of dwarf galaxies; after all, the Local Group represents only a mild
overdensity on a mass scale close to the characteristic clustering mass today. Cosmic web stripping is thus very likely a crucial ingredient of the star formation history of dwarf galaxies so far missing from models that rely solely or largely on the assembly of the dark matter.

Its appeal comes in part from its strong dependence on halo mass. Because of the $(V_{\rm p}/V_{200})^2$ scaling mentioned above, a pancake sweeping past halos of different mass could fully strip the gas reservoir of low-mass halos whilst leaving massive halos relatively undisturbed. This might help to explain the steep galaxy
mass-halo mass relation required to match the faint-end of the galaxy luminosity function. We emphasize that stripping must operate in concert with cosmic reionization and feedback if $\Lambda$CDM models are to be viable. This is because Mpc-scale pancakes and filaments do not develop typically until $z\sim 2$, so a further mechanism that prevents most baryons in low-mass halos from turning into stars before then is still needed.

How can we test these ideas observationally?  Because cosmic web stripping affects only baryons, its effects are intertwined with those of radiative cooling, reionization, star formation, and feedback, and it is difficult to devise a clean and conclusive test. It is nevertheless encouraging that several observations are in natural accord with a scenario where cosmic web stripping plays a substantial role. The existence of isolated dwarf spheroidal galaxies devoid of gas and made up of mostly old stars, such as Cetus and Tucana, is one such observation. Indeed, cosmic web stripping might explain why these galaxies ceased forming stars much later than the epoch of reionization~\citep{Monelli2010a,Monelli2010b}, the oft-cited culprit of halting star formation in dwarf galaxies. 

Cosmic web stripping should also leave an environmental imprint, as galaxies more severely affected by interaction with the cosmic web should cluster differently from others that are impacted less. It is not possible to divine what the actual imprint might be from one Local Group simulation, but this is an area where future simulation and observational work should prove especially fruitful. Detection of the telltale signs of ram-pressure stripping in isolated dwarfs would be especially interesting, and it might have already been observed~\citep{McConnachie2007}.  Finally, we note that a galaxy might interact with the cosmic web several times during its evolution, and that each of these events might serve to shape different episodes of star formation. It is certainly tempting to speculate that the existence of multiple populations in dwarf galaxies like Carina~\citep{Monelli2003,Bono2010} might originate from its travels through the cosmic web. Targeted future work will help us to gauge the true importance of 
the cosmic web in shaping the formation and evolution of the faintest galaxies.

\section{Acknowledgments}

The simulations were performed at Leibniz Rechenzentrum Munich (LRZ) and at Barcelona Supercomputing Center
(BSC). GY acknowledges support from MINECO (Spain) through grants AYA2009-13875-C03-02 and FPA2009-08958. YH has been partially supported by the Israel Science Foundation (13/08).

\end{document}